\documentstyle{article}
\begin{document}

\author{Gunn A. Quznetsov \\
quznets@geocities.com}
\title{THE PHYSICS SPACE DIMENSION}
\maketitle

\begin{abstract}
All fermions and all interactions between fermions are expressed by the
Cayley numbers in our space-time.

\end{abstract}

\section{3+1 SPACE-TIME}

Let $A(t,\overrightarrow{x})$ be the event, which can be expressed as: ''The
particle $e_A$ is detected in the space point $\overrightarrow{x}$ at the
time moment $t$'' and $B(t_0,\overrightarrow{x_0})$ - as:

''The particle $e_B$ is detected in the space point $\overrightarrow{x_0}$
at the time moment $t_0$''.

Let $\rho (t,\overrightarrow{x})$ be the probability density of the event $A$%
. That is

\[
\int \int \int_{\left( V\right) }d\overrightarrow{x}\cdot \rho (t,%
\overrightarrow{x}) 
\]
equals to the probability to detect the particle $e_A$ in the space domain $%
V $ at the time moment $t$.

Let $\rho _c(t,\overrightarrow{x}|t_0,\overrightarrow{x_0})$ be the
conditional probability density of the event $A$ for the event $B$. That is

\[
\int \int \int_{\left( V\right) }d\overrightarrow{x}\cdot \rho (t,%
\overrightarrow{x}|t_0,\overrightarrow{x_0}) 
\]
equals to the probability to detect the particle $e_A$ in the space domain $%
V $ at the time moment $t$, if the particle $e_B$ is detected in the space
point $\overrightarrow{x_0}$ at the time moment $t_0$.

In this case, if

\[
\rho _c(t,\overrightarrow{x}|t_0,\overrightarrow{x_0})=g(t,\overrightarrow{x}%
|t_0,\overrightarrow{x_0})\cdot \rho (t,\overrightarrow{x})\mbox{,} 
\]
then the function $g(t,\overrightarrow{x}|t_0,\overrightarrow{x_0})$ is the
interaction function for $e_A$ and $e_B$. If $g(t,\overrightarrow{x}|t_0,%
\overrightarrow{x_0})$ $=1$ then the particles $e_A$ and $e_B$ do not
interact.

In the Quantum Theory a probability density equals to the quadrate of the
state vector module. A fermion state vector is the 4-component complex
vector. Hence, a fermion state vector has got 8 real components. Therefore,
some conformity between such vectors and the octaves (the Cayley numbers)
can be determined.

Let $\rho =\Psi ^{\dagger }\cdot \Psi $ and $\rho _c=\Psi _c^{\dagger }\cdot
\Psi _c$. Here: $\Psi $ and $\Psi _c$ are the 4-component complex state
vectors. And $\Psi $ and $\Psi _c$ are the octaves by this conformity.

Because the Cayley algebra is the division algebra, then the octave $\varphi 
$ exists, for which: $\Psi _c=\varphi \bullet \Psi $ (here $\bullet $ is the
symbol of the algebra Cayley product).

Because the Cayley algebra is the normalized algebra, then $g=\varphi
^{\dagger }\cdot \varphi $.

Therefore, all fermion interactions can be expressed by the octaves in the $%
3+1$ space-time.

\section{$\mu $+1 SPACE-TIME}

Let us consider the probability density $\rho \left( t,\overrightarrow{x}%
\right) \cite{LNL}$ for some point-event $A\left( t,\overrightarrow{x}%
\right) $ in the $\mu +1$ space-time. That is

\[
\int_{\left( V\right) }d\overrightarrow{x}\cdot \rho \left( t,%
\overrightarrow{x}\right) 
\]

is the probability for $A$ to happen in the space domain $\left( V\right) $
at the time moment $t$ in the $\mu +1$ space-time.

Let $\overrightarrow{j}\left( t,\overrightarrow{x}\right) $ be the
probability current vector \cite{LNL2}. In this case 

\[
\left\langle \rho \left( t,\overrightarrow{x}\right) ,\overrightarrow{j}%
\left( t,\overrightarrow{x}\right) \right\rangle 
\]

is the probability density $\mu +1$vector.

The Clifford set of the range $s$ is the set $K$ of the $s\times s$ complex
matrices for which:

1) if $\gamma \in K$ then $\gamma ^2=1$ (here $1$ is the identity $s\times s 
$ matrix);

2) if $\gamma \in K$ and $\beta \in K$ then $\gamma \cdot \beta +\beta \cdot
\gamma =0$ (here $0$ is the zero $s\times s$ matrix);

3) There does not exist the $s\times s$ matrix $\zeta $ which anticommutates
with all $K$ elements, for which $\zeta ^2=1$, and which is not the element
of $K$.

For example, the Clifford pentad $\left\langle \beta ^1,\beta ^2,\beta
^3,\beta ^4,\gamma ^0\right\rangle $ \cite{LNL3} is the Clifford set of the
range $4$.

By \cite{ZH} for every natural number $z$ the Clifford set of the range $2^z$
exists.

For every probability density vector $\left\langle \rho \left( t,%
\overrightarrow{x}\right) ,\overrightarrow{j}\left( t,\overrightarrow{x}%
\right) \right\rangle $ the natural number $s$, the Clifford set $K$ of
range $s$ and the complex $s$-vector $\Psi \left( t,\overrightarrow{x}%
\right) $exist, for which: $\gamma _n\in K$ and

\begin{equation}  \label{1}
\Psi \left( t,\overrightarrow{x}\right) ^{\dagger }\cdot \Psi \left( t,%
\overrightarrow{x}\right) =\rho \left( t,\overrightarrow{x}\right) ,
\end{equation}

\begin{equation}  \label{1'}
\Psi \left( t,\overrightarrow{x}\right) ^{\dagger }\cdot \gamma _n\cdot \Psi
\left( t,\overrightarrow{x}\right) =j_n\left( t,\overrightarrow{x}\right) .
\end{equation}

In this case let $\Psi \left( t,\overrightarrow{x}\right) $ be named as the $%
s$-spinor for $\left\langle \rho \left( t,\overrightarrow{x}\right) ,%
\overrightarrow{j}\left( t,\overrightarrow{x}\right) \right\rangle $.

If $\left\langle \rho \left( t,\overrightarrow{x}\right) ,\overrightarrow{j}%
\left( t,\overrightarrow{x}\right) \right\rangle $ obeys to the continuity
equation \cite{LNL4} then from (\ref{1}), (\ref{1'}): $\Psi \left( t,%
\overrightarrow{x}\right) $ is fulfilled to the Dirac equation
generalization on the $\mu +1$ space-time.

Let $\rho _c(t,\overrightarrow{x}|t_0,\overrightarrow{x_0})$ be the
conditional probability density of the event $A$ for the event $B$, but in
the $\mu +1$ space-time, too. And if

\[
\rho _c(t,\overrightarrow{x}|t_0,\overrightarrow{x_0})=g(t,\overrightarrow{x}%
|t_0,\overrightarrow{x_0})\cdot \rho (t,\overrightarrow{x})\mbox{,} 
\]
then the function $g(t,\overrightarrow{x}|t_0,\overrightarrow{x_0})$ is the
interaction function for $e_A$ and $e_B$, too, but in the $\mu +1$
space-time.

Let $\Psi _c$ and $\Psi $ be the $s$-spinors, for which: $\rho =\Psi
^{\dagger }\cdot \Psi $ and $\rho _c=\Psi _c^{\dagger }\cdot \Psi _c$.

Let $\Psi _c$ and $\Psi $ are the elements of the algebra $\Im $ with the
product $*$, and for every $\Psi _c$ and $\Psi $ the element $\varphi $ of $%
\Im $ exists, for which: $\Psi _c=\varphi *\Psi $ and $\varphi ^{\dagger
}\cdot \varphi =g$.

In this case $\Im $ is the division normalized algebra and the $\Im $
dimension is not more than 8 from the Hurwitz theorem \cite{O1} (The every
normalized algebra with the unit is isomorphic to alone from the followings:
the real numbers algebra $R$, the quaternions algebra $K$, or the octaves
algebra $\acute O$) and from the generalized Frobenius theorem \cite{O2}
(The division algebras have got the dimension 1,2,4 or 8, only). Therefore,
the Clifford set matrices size are not more than $4\times 4$ (the Clifford
matrices are the complex matrices). Such Clifford set contains not more than
5 elements. The diagonal elements of this pentad defines the space, in which
the physics particle move. This space dimension is not more than 3 \cite
{LNL5}. Hence, in this case we have got 3+1 space.

If $\mu >3$ then for every algebra the interaction functions exist, which do
not belong to this algebra. I'm name such interactions as the supernatural
for this algebra interactions.

\section{RESUME}

1. All fermions and all interactions between fermions are expressed by the
octaves in our space-time.

2. The probability, which is defined by the relativistic $\mu +1$-vector of
the probability density, fulfils to the Quantum Theory principles.

3. In the $\mu $+1 space-time: if $\mu \leq 3$ then the supernatural
interactions do not happen, if $\mu >3$ then the supernatural interactions
happen.

\section{APPENDIX. CAYLEY ALGEBRA}

The Cayley algebra \'O has got basis on the 8 dimtnsional real linear space.
The orthogonal normalized basic elements of \'O are: $1$, $i$, $j$, $k$, $E$%
, $I$, $J$, $K$. The product of \'O is defined by the following rules ($%
\grave a\bullet \grave e$):

$\left\| 
\begin{array}{cc}
\grave a\backslash \grave e & 
\begin{array}{cccccccc}
.1 & ...i & ..j.. & ..k.. & .E. & .I. & .J.. & .K
\end{array}
\\ 
\begin{array}{c}
1 \\ 
i \\ 
j \\ 
k \\ 
E \\ 
I \\ 
J \\ 
K
\end{array}
& 
\begin{array}{cccccccc}
1 & i & j & k & E & I & J & K \\ 
i & -1 & k & -j & I & -E & -K & J \\ 
j & -k & -1 & i & J & K & -E & -I \\ 
k & j & -i & -1 & K & -J & I & -E \\ 
E & -I & -J & -K & -1 & i & j & k \\ 
I & E & -K & J & -i & -1 & -k & j \\ 
J & K & E & -I & -j & k & -1 & -i \\ 
K & -J & I & E & -k & -j & i & -1
\end{array}
\end{array}
\right\| $

\end{document}